\documentclass[12pt]{article}
\usepackage{amssymb,amsmath,epsfig}
\usepackage{cite}
\allowdisplaybreaks

\begin{document}
\title{\bf On the Role of $f(G,T)$ Terms in Structure Scalars}
\author{Z. Yousaf \thanks{zeeshan.math@pu.edu.pk}\\
Department of Mathematics, University of the
Punjab,\\ Quaid-i-Azam Campus, Lahore-54590, Pakistan}
\date{}
\maketitle
\begin{abstract}
This work is devoted to explore the effects of $f(G,T)$ terms on the
study of structure scalars and their influences in the formulations
of the Raychaudhuri, shear and Weyl scalar equations. For this
purpose, we have assumed non-static spherically symmetric geometry
coupled with shearing viscous locally anisotropic dissipative matter
content. We have developed relations among the Misner-Sharp mass,
Weyl scalar, matter and structure variables. We have also formulated
set of $f(G,T)$ structure scalars after orthogonally breaking down
of the Riemann curvature tensor. The influences of these scalar
functions in the modeling of relativistic radiating spheres are also
studied. The factor involved in the emergence of inhomogeneities is
also explored for the constant and varying modified curvature
corrections. We inferred that $f(G,T)$ structure scalars could lead
provide an effective tool to study Penrose-Hawking singularity
theorems and Newman-Penrose formalism.
\end{abstract}

\section{Introduction}

Even after the introduction of one century, the well-known General
Relativity (GR) is considered itself as the most effective
relativistic theory to investigate the gravitational interactions in
the sel-gravitating system and the related scenarios. However, the
need for the exploration of its alternatives can not be ignored that
stem from the confrontation that GR presently experience \cite{qs1,qs2,qs3,qs4,qs5,qs6,qs7}.
The current burning issues of dark matter (DM) and accelerated expansion of our cosmos (along with the
cosmological constant ($\Lambda$) problem) urged the need of hour
towards GR modification. After performing a detailed analysis on the
modified relativistic dynamics, Qadir et al. \cite{zs3} suggested that
GR may need to extend for the discussion of DM problem, quantum
gravity effects and some other related burning issues. Many
theoretical scheme have been seen in the literature with the aim to
handle such problems.

The modified gravity theories (MGTs) are found to be one of the
attractive mathematical tools to explore the dark sector of the
universe. various DE theoretical models are being suggested by
modifying the gravitational part of the usual Einstein-Hilbert (EH)
action (for details, please see \cite{v41, b2a, R-DE-MG,martin1}).
Nojiri and Odintsov \cite{ya3} found some observationally
well-consistent modified gravity models (MGM) for the study of dark
cosmic aspect. They inferred that there could be few very
interesting $f(R)$ models that can provide the dynamical study of
relativistic systems compatible with the solar systems tests. Yousaf
and Bhatti \cite{mnras1} claimed that some MGM prefer to host
super-massive but with relatively smaller radii compact objects.

The notable $f(R)$ theory is found to be suffer from some scale
dependence effects, which can not be ignored for the the exploration
of one post-Newtonian term alone. Furthermore, it could be
interesting to include amalgams of curvature measuring mathematical
quantities, like Riemann tensor ($R_{\alpha\beta\mu\nu}$), Ricci
tensor ($R_{\mu\nu}$) and its scalar ($R$) in the EH action. This
approach results with the $f(G)$ gravity theory, in which $G$ is a
topological invariant Gauss-Bonnet term and is defined as $G = R -
4{R_{\mu \nu }}{R^{\mu \nu }} + {R_{\mu \nu \alpha \beta }}{R^{\mu
\nu \alpha \beta }}$. This theory was first introduced in
\cite{martin3}. Such theories could help out to understand cosmic
inflationary epochs and to see the transition phases of our cosmos
in the accelerating phase from the corresponding deceleration era.
It could be considered as a viable substitute to understand DE
issues \cite{g11,g12,martin1}. Recently, this
theory has been modified further by including corrections from the
trace of energy-momentum tensor ($T$) in the EH action. This
gravitational model is known as $f(G,T)$ theory. Just like as Harko
\textit{et al.} \cite{ya9} modified $f(R)$ to $f(R,T)$ theory.

Houndjo \cite{ya10} calculated few observationally viable
mathematical models for $f(R,T)$ gravity and claimed that these
could encode our cosmic dynamics associated with the matter
dominated regime. Baffou \textit{et al.} \cite{ya13} applied
perturbation theory to analyze the behavior of some cosmic
mathematical models in the background of power law and de-Sitter
spacetime. Bamba \emph{et al.} \cite{z5d} studied the influences of
extra degrees of freedom mediated by modified gravity terms on the
accelerating nature of our expanding universe. The phenomenon of
gravitational instabilities for the self-gravitating stellar
interiors were examined in the mode of $f(R)$ gravity by
\cite{mart5,mart6,mart7}. Yousaf and his collaborators studied the
collapse rate of the compact bodies in account of various modified
theories with the planar \cite{7p1,7p2,7p3}, spherical \cite{7s1,7s2,7s3,7s4,7s5,7s6} and
cylindrical \cite{7c1,7c3} environments. Ilyas et al. studied the impact
of quadratic and exponential $f(R,T)$ models in the mathematical
modeling and stability of observationally well-consistent spherical
stars. Moraes \emph{et al.} \cite{9} explored hydrostatic state of
strange stars in order to investigate their stable regimes with
$f(R,T)=R+2\lambda T$ model. Recently, Bhatti \emph{et al.} \cite{gt1} performed
computational analysis to check the role of logarithmic $f(G,T)$
models on the existence of Her X-1 compact star.

Herrera \emph{et al.} \cite{12,13} studied few dynamical features of
cylindrical as well as spherical gravitational collapse after
evaluating junction conditions. Tewari \emph{et al.} \cite{15}
examined effects of locally anisotropic pressure on the spherical
collapse. Sharif and Yousaf \cite{16rt,16r} analyzed the problem of
collapsing system by investigating the role of matter variables in
modified gravity. Recently, Yousaf \emph{et al.} \cite{epjc1,cqg}
smoothly matched non-static irrotational cylindrical spacetime with
an exterior geometry of Einstein-Rosen bridge and examined the issue
of dynamical instability. Sahoo \emph{et al.} \cite{sah1} considered the problem of cosmic evolution and
discussed some kinematical features of the temporal varying deceleration parameters. Recently, Moraes \emph{et al.} \cite{sah2a,sah2b,sah2c}
studied various interesting cosmic and stellar issues in the field of $f(R,T)$ theory.

The self-gravitating relativistic system would undergo the
collapsing phase, once it experiences inhomogeneities in its energy
density. This has urged many researchers to explore those factors
that are involved in the emergence of irregularities in the
initially homogeneous celestial object. Penrose and Hawking \cite{ya26}
discussed the causes of inhomogeneous energy density (IED) with the help of tidal forces producing tensor called
the Weyl tensor for the relativistic sphere. Herrera \textit{et al.} \cite{ya27} explored few
parameters involved in the maintenance of IED in an environment of anisotropic
spheres and inferred that effects of anisotropy in the stellar pressure could to the formations
of naked singularity (NS). In the mathematical viewpoint, Virbhadra \emph{et al.}
\cite{vir1a,vir1b,vir2} provided formula for helping the relativistic to analyze the formation of NS and black holes during
the evolutionary phases of stars.

Herrera \textit{et al.} \cite{ya29} studied the gravitational time
arrow in the Einstein gravity for the radiating compact stars and
presented a relation among the tidal forces, IED and locally
anisotropic pressure. Herrera \textit{et al.} \cite{ya30} explored
the variations of expansion scalar in the maintenance of IED foe the
viscous charged spherical objects. Yousaf \emph{et al.} \cite{y1t}
extended these results for the case of radiating relativistic sphere
with extra degrees of freedom coming from modified gravity. They
concluded that modified gravity terms has greatly influence the role
of Weyl scalar in the maintenance of IED. Bhatti and his
collaborators \cite{b1ta,b1tb} examined some characteristics of collapsing
spheres and explored the corresponding IED factors in modified
gravity. Herrera \emph{et al.} \cite{ltb} and Herrera
\cite{entropy1} evaluated transport equations for the spherically
symmetric matter distributions which undergoes in the collapsing
state as seen by a tilted observer. Yousaf \emph{et al.} \cite{y2t}
modified these results and looked into the effects of Palatini
$f(R)$ terms in the rate of gravitational collapse. Most recently,
Herrera \cite{jpc} investigated that why observations of
non-comoving congruences observe dissipation from the celestial
objects which seem to be isentropic for moving observers.

Here, we have extended the work of Herrera \emph{et al.}
\cite{ya30} with the aim to analyze the influences of $f(G,T)=\alpha
G^n(\beta G^m+1)+\lambda T$ model in the mathematical modeling of
structure scalars, Weyl, expansion and shear equations. We shall
outline this paper as under. The coming section will describe
some equations required to describe $f(G,T)$ gravity and spherical
dissipative viscous matter configurations. Section \textbf{3} is
devoted to evaluate modified $f(G,T)$ structure scalars obtained
from the orthogonal breaking down of the Riemann curvature tensor
with power law Gauss Bonnet and linear $T$ terms. The role of
differential equation corresponding to Weyl, shear and expansion
scalars are also discussed in this gravity. In section \textbf{4},
we demonstrate the working of $f(G,T)$ scalar functions in the
evolution of IED over the surface of the initially smooth dust ball
with constant $G$ and $T$ terms. Finally, we describe our conclusion
in the last section.

\section{Spherical Viscous Spherical System and $f(G,T)$ Gravity}

The action function for the theory of $f(G,T)$ gravity can be written as
\begin{equation}\label{action1}
S={\frac{1}{\kappa^2}}\int{{d^4}x\sqrt{-g}\left[{\frac{R}{2}+
f(G,T)}\right]+{S_M}\left({{g^{\mu\nu}},\psi}\right)},
\end{equation}
where ${\kappa ^2}$ is a coupling constant which is taken to be unity here. In the above equation,
$GT$ is the Gauss-Bonnet term whose expression with the help of the Ricci tensor $(R_{\mu\nu})$, scalar ($R$)
and the Riemann curvature tensors $({R_{\mu\nu\alpha\beta}})$ can be given as  $$G={R_{\mu\nu\alpha\beta
}}{R^{\mu\nu\alpha\beta}}-4{R_{\mu\nu}}{R^{\mu\nu}}+R.$$ Further, $g$ and $S_M$ are the metric tensor determinant
and the matter action, respectively. The quantity $T$ is the trace of the following energy momentum tensor
\begin{equation}\label{action2}
T_{\gamma\delta}=-\frac{2}{\sqrt{-g}}\frac{\delta(\sqrt{-g}L_m)}{\delta g^{\gamma\delta}}.
\end{equation}
This equation after some manipulations and assumptions, can be rewritten as
\begin{equation}\label{action3}
T_{\gamma\delta}=g_{\gamma\delta}L_m-\frac{2\partial L_m}{\partial g_{\gamma\delta}},
\end{equation}
whose $\delta$ variations gives
\begin{equation}\label{action4}
\frac{\delta T_{\gamma\delta}}{\delta g^{\mu\nu}}=\frac{\delta g_{\gamma\delta}}{\delta g^{\mu\nu}}L_m -\frac{2\partial^2L_m}{\partial g^{\mu\nu}\partial g^{\gamma\delta}}+g_{\gamma\delta}\frac{\partial L_m}{\partial g^{\mu\nu}}.
\end{equation}
Upon varying Eq.(\ref{action1}) with the metric tenor, we obtain the following $f(G,T)$ equations of motion
\begin{equation}\label{field equation}
{G_{\gamma\delta}}=T_{\gamma\delta}^{eff},
\end{equation}
where ${G_{\gamma\delta }={R_{\gamma\delta}}-{\frac{1}{2}}R{g_{\gamma\delta}}}$ and
$T_{\gamma\delta}^{eff}$ is
\begin{align}\nonumber
T_{\gamma\delta}^{eff}&={\kappa^2}{T_{\gamma\delta}}-({T_{\gamma\delta
}} +{\Theta _{\gamma\delta}}){f_T}(G,T)+\frac{1}{2}{g_{\gamma\delta
}}f(G,T)-(2R{R_{\gamma\delta}}-4R_\gamma^\varepsilon{R_{\varepsilon
\beta }}\\\nonumber &-4{R_{\gamma\varepsilon\delta\eta
}}{R^{\varepsilon\eta }} +2R_\gamma ^{\varepsilon \eta \mu
}{R_{\delta \varepsilon \eta \mu }}){f_G}(G,T)-(2R{g_{\gamma\delta
}}{\nabla ^2} - 2R{\nabla _\gamma }{\nabla _\delta }- 4{R_{\gamma
\delta }}{\nabla ^2}\\\nonumber & -4{g_{\gamma\delta
}}{R^{\varepsilon \eta }}{\nabla _\varepsilon }{\nabla _\eta }
+4R_\gamma ^\varepsilon {\nabla _\delta }{\nabla _\varepsilon }+
4R_\delta ^\varepsilon {\nabla _\gamma }{\nabla _\varepsilon }+
4{R_{\gamma \varepsilon \delta \eta }}{\nabla ^\varepsilon }{\nabla
^\eta }){f_G}(G,T),
\end{align}
where $\nabla_\gamma$ is an operator for covariant derivations and ${\nabla ^2}\equiv
{\nabla _\gamma }{\nabla ^\gamma }$ and the the subscript in the above terms describe the respective partial differentiations.
Moreover, the expression for $\Theta_{\gamma\delta}$ is given by
\begin{align}\label{action5}
\Theta_{\gamma\delta}=g^{\mu\nu}\frac{\delta T_{\mu\nu}}{\delta g_{\gamma\delta}}.
\end{align}
After using Eq.(\ref{action4}), Eq.(\ref{action5}) turns out to be
\begin{align}\nonumber
\Theta_{\gamma\delta}=g_{\gamma\delta}L_m-2T_{\gamma\delta}-2g^{\mu\nu}\frac{\partial^2 L_m}{\partial g^{\gamma\delta}\partial g^{\mu\nu}}.
\end{align}
The trace of Eq.(\ref{field equation}) is given by
\begin{align}\nonumber
T+R-(\Theta+T)f_T+2Gf_G+2f-2R\nabla^2f_G+4R_{\gamma\delta}\nabla^\gamma\nabla^\delta f_G=0.
\end{align}
The one of the aims of this work is to analyze the role of heat dissipation $(q_\gamma)$, radiation density $(\varepsilon)$ and pressure anisotropicity $\Pi\equiv P_r-P_\bot$ in the definitions of modified scalars functions, the scalars whose expressions can be achieved by the orthogonal breaking down of the
Riemann tensor. For this purpose, We assume the following form of the stress-energy tensor as
\begin{equation}\label{4}
T_{\lambda\nu}={\mu}V_\lambda V_\nu+P_{\bot}h_{\lambda\nu}+\Pi
\chi_\lambda\chi_\nu-2{\eta}{\sigma}_{\lambda\nu}+{\varepsilon}l_\lambda l_\nu+q(\chi_\nu
V_\lambda+\chi_\lambda V_\nu),
\end{equation}
where $\eta$ describes the magnitude of the shear tensor $\sigma_{\gamma\delta}$ and $h_{\gamma\delta}$ is the projection
tensor that can be given via four vector $V_\gamma$ as $h_{\gamma\delta}=g_{\gamma\delta}+V_{\gamma}V_{\delta}$. Further,
$l^\beta$ and $\chi^\beta$ are the null and radial four vectors,
respectively.

The extra curvature corrections of the $f(G,T)$ gravity can be
invoked by considering separate formulations for the functions of
$G$ and $T$. Therefore, we choose $f(G,T)$ model of the type
\begin{align}\label{generalmodel}
f(G,T)=f(G)+g(T).
\end{align}
The models of such types could be regarded as the possible corrections in the well-known $f(G)$ gravity.
Nojiri and Odintsov firstly introduced $f(G)$ gravity in \cite{martin3}. The choices of models given above could be considered as a possible toy models for the understanding of the dark sector of the universe. Here, we use linear $g(T)$
with the aim to see some striking consequences on the dynamics of spherical stars on the
basis of extra curvature terms stem from the $f(G)$ gravity. Therefore, we have
\begin{align*}
f(G,T)=f(G)+\lambda T,
\end{align*}
in which $\lambda$ is a constant number. To include Gauss-Bonnet corrections, we consider $f(G)$
model containing three different constants $\alpha,~\beta$ and $m$, given as follows \cite{g16}
\begin{equation}\label{model2}
f(G)=\alpha  G^n \left(\beta  G^m+1\right),
\end{equation}
where $n>0$. This model was introduced to understand the finite time future singularities.

Now, we consider a diagonally symmetric non-static general form of spherical spacetime as
\begin{equation}\label{3}
ds^2=B^2(t,r)dr^{2}-A^2(t,r)dt^{2}+C^2(d\theta^{2}
+\sin^2\theta{d\phi^2}),
\end{equation}
where the scale factors $A,~B$ and $C$ are considered to be positive. We assume that
our relativistic sphere have an anisotropic shearing viscous and radiating interior whose mathematical form is mentioned in
Eq.(\ref{4}). In the non-tilted frame of reference, the four vectors appearing in the formulation
of usual energy momentum tensor (\ref{4}) are obeying the relations
\begin{eqnarray*}
&&\chi^{\nu}\chi_{\nu}=1,\quad V^{\nu}V_{\nu}=-1,
\quad\chi^{\nu}V_{\nu}=0,\\\nonumber
&&l^\nu V_\nu=-1, \quad V^\nu q_\nu=0, \quad l^\nu
l_\nu=0,
\end{eqnarray*}
along with their definitions
$$V^{\nu}=\frac{1}{A}\delta^{\nu}_{0},~
\chi^{\nu}=\frac{1}{C}\delta^{\nu}_{1},~
l^\nu=\frac{1}{A}\delta^{\nu}_{0}+\frac{1}{B}\delta^{\nu}_{1},~
q^\nu=q(t,r)\chi^{\nu}.$$

The corresponding expansion scalar and shear tensor are given by
\begin{equation*}
\sigma A=\left(\frac{\dot{H}}{H}
-\frac{\dot{C}}{C}\right),\quad \Theta_1 A=\left(\frac{\dot{H}}{H}+\frac{2\dot{C}}{C}\right).
\end{equation*}
where dot notation is representing $\frac{\partial}{\partial t}$ operator. The $f(G,T)$ field equations for
Eqs.(\ref{4})-(\ref{3}) are
\begin{align}\label{6}
G_{00}&={A^2}\left[{\mu}+{\varepsilon}+\varepsilon\lambda
-\frac{\alpha}{2}\left\{\beta(1-n-m)G^m+(1-n)\right\}G^n-\frac{\lambda T}{2}+\frac{\varphi_{00}}{A^2}
\right],\\\label{7} G_{01}&={BA}\left[-{(1+\lambda)}(q+{\varepsilon})
+\frac{\varphi_{01}}{BA}\right],\\\label{8}
G_{11}&={B^2}\left[\mu
\lambda+(1+\lambda)(P_r+\varepsilon-\frac{4}{3}\eta{\sigma})
+\frac{\alpha}{2}\left\{\beta(1-n-m)G^m+(1-n)\right\}G^n+\frac{\lambda T}{2}+\frac{\varphi_{11}}{H^2}\right],\\\label{9}
G_{22}&={C^2}\left[(1+\lambda)({P_{\bot}}+\frac{2}{3}\eta{\sigma})+\mu
\lambda
+\frac{\alpha}{2}\left\{\beta(1-n-m)G^m+(1-n)\right\}G^n+\frac{\lambda T}{2}+\frac{\varphi_{22}}{C^2}\right],
\end{align}
where the expressions of $G_{\gamma\delta}$ can be found from \cite{ya30}. Here, the notation prime
indicates radial partial differentiation. Now, we consider the definitions of the fluid
4-velocity as
\begin{eqnarray}\label{10}
U=D_{T}C=\frac{\dot{C}}{A}.
\end{eqnarray}
The mass $\mathfrak{m}$ for the spherical structures can be calculated via Misner-Sharp directions as \cite{ya32}
\begin{equation}\label{11}
\mathfrak{m}(t,r)=\frac{C}{2}\left(1+\frac{\dot{C}^2}{A^2}
-\frac{C'^2}{H^2}\right).
\end{equation}
The variations in the physical quantity $\mathfrak{m}$ can found through Eqs.(\ref{6})-(\ref{8}) and (\ref{10}) as
\begin{align}\label{12}
D_T{\mathfrak{m}}&=\frac{-1}{2}\left[U\left\{(1+\lambda)(\bar{P}_r-\frac{4}{3}\eta\sigma)+\lambda\mu
+\frac{\alpha}{2}\left\{\beta(1-n-m)G^m+(1-n)\right\}G^n+\frac{\lambda T}{2}\right.\right.\\\nonumber
&\left.\left.+\frac{\varphi_{11}}{H^2}\right\}+E
\left\{{(1+\lambda)}\bar{q}-\frac{\varphi_{01}}{BA}\right\}\right],\\\nonumber
D_C\mathfrak{m}&=\frac{C^2}{2}\left[\bar{\mu}+\lambda\varepsilon
-\frac{\alpha}{2}\left\{\beta(1-n-m)G^m+(1-n)\right\}G^n-\frac{\lambda T}{2}
+\frac{\varphi_{00}}{A^2}-\frac{U}{E}\left\{\frac{\varphi_{01}}{AH}\right.\right.\\\label{13}
&\left.\left.-{(1+\lambda)}\bar{q}\right\}\right],
\end{align}
where $\bar{H}=h+\varepsilon,$ while
$D_{T}=\frac{1}{A} \frac{\partial}{\partial t}$. Equation (\ref{13}) can be remanipulated through $E\equiv \frac{C'}{H}$ as
\begin{align}\nonumber
\mathfrak{m}&=\frac{1}{2}\int^C_{0}{C^2}\left[\bar{\mu}+\lambda\varepsilon-\frac{\alpha}{2}\left\{\beta(1-n-m)G^m+(1-n)\right\}G^n-\frac{\lambda T}{2}
+\frac{\varphi_{00}}{A^2}+\frac{U}{E}\left\{\frac{\varphi_{01}}{BA}\right.\right.\\\label{14}
&\left.\left.+{(1+\lambda )}\bar{q}\right\}\right]dC.
\end{align}
The influences of tidal forces can also be described as
\begin{eqnarray}\label{15}
E\equiv\frac{C'}{H}=\left[1+U^{2}-\frac{2\mathfrak{m}(t,r)}{C}\right]^{1/2}.
\end{eqnarray}
Equations (\ref{12})-(\ref{15}) yield
\begin{align}\nonumber
\frac{3\mathfrak{m}}{C^3}&=\frac{3\kappa}{2C^3}\int^r_{0}\left[\bar{\mu}+\lambda\varepsilon
-\frac{\alpha}{2}\left\{\beta(1-n-m)G^m+(1-n)\right\}G^n-\frac{\lambda T}{2}
+\frac{\varphi_{00}}{A^2}+\frac{U}{E}\left\{{(1+\lambda)}\bar{q}\right.\right.\\\label{16} &\left.\left.
+\frac{\varphi_{01}}{BA}\right\}C^2C'\right]dr,
\end{align}
This expression has connected $f(G,T)$ dark source corrections with the dissipative fluid variables, like spherical mass,
energy density, heat conduction. The decomposition of the Weyl
tensor provide us two major parts namely magnetic
and electric denoted respectively by $H_{\alpha\beta}$ and $E_{\alpha\beta}$.
These two are defined respectively as
\begin{equation*}
H_{\alpha\beta}=\frac{1}{2}\epsilon_{\alpha \gamma
\eta\delta}C^{\eta\delta}_{~~\beta{\rho}}V^\gamma
V^{\rho}=\tilde{C}_{\alpha
\gamma\beta\delta}V^{\gamma}V^\delta=,\quad
E_{\alpha\beta}=C_{\alpha\phi\beta \varphi}V^{\phi}V^{\varphi},
\end{equation*}
where $\epsilon_{\lambda\mu\nu\omega}\equiv\sqrt{-g}\eta_{\lambda\mu\nu\omega}$
with $\eta_{\lambda\mu\nu\omega}$ as a Levi-Civita symbol. The quantity $E_{\lambda\nu}$ can be given
via $V_\gamma$ as
\begin{equation*}\nonumber
E_{\lambda\nu}=\left[\chi_{\lambda}\chi_{\nu}-\frac{g_{\lambda\nu}}{3}
-\frac{1}{3}V_\lambda V_\nu\right]\mathcal{E},
\end{equation*}
where $\mathcal{E}$ is a Weyl scalar whose expression can be given alternatively as
\begin{eqnarray}\nonumber
\mathcal{E}&=&\left[
\left(\frac{\dot{A}}{A}+\frac{\dot{C}}{C}\right)\left(\frac{\dot{B}}{B}-\frac{\dot{C}}{C}\right)-\frac{\ddot{B}}{B}+\frac{\ddot{C}}{C}\right]
\frac{1}{2A^{2}}-\frac{1}{2C^{2}}\\\label{17}
&-&\left[-\left(\frac{A'}{A}-\frac{C'}{C}\right)\left(\frac{C'}{C}+\frac{B'}{B}\right)
+\frac{C''}{C}-\frac{A''}{A}\right]\frac{1}{2B^{2}}.
\end{eqnarray}
The Weyl scalar $\mathcal{E}$ with extra degrees of freedom mediated from $f(G,T)$
gravity is given by
\begin{align}\nonumber
\mathcal{E}&=\frac{1}{2}\left[\bar{\mu}+\lambda\varepsilon-(1+\lambda)(\bar{\Pi}-2\eta\sigma)
-\frac{\alpha}{2}\left\{\beta(1-n-m)G^m+(1-n)\right\}G^n-\frac{\lambda T}{2}
-\frac{\varphi_{00}}{A^2}\right.\\\nonumber
&\left.+\frac{\varphi_{11}}{B^2}
-\frac{\varphi_{22}}{C^2}\right]-\frac{3}{2C^3}
\int^r_{0}{C^2}
\left[\bar{\mu}+\lambda\varepsilon-\frac{\alpha}{2}\left\{\beta(1-n-m)G^m+(1-n)\right\}G^n-\frac{\lambda T}{2}
+\frac{\varphi_{00}}{A^2}\right.\\\label{18}
&\left.+\frac{U}{E}\left\{(1+\lambda)\bar{q}
-\frac{\varphi_{01}}{BA}\right\}\right]C'dr,
\end{align}
where $\bar{\Pi}\equiv\bar{P}_r-P_\perp$.

\section{Structure Scalars and $f(G,T)$ Gravity}

This section discusses the analytical computation of extended forms of structure scalars
for the spherical relativistic interiors framed with $f(G)+\lambda T$ gravity. In this background,
we define couple of tensorial expressions, $X_{\alpha\beta}$ and
$Y_{\alpha\beta}$, that were presented firstly by Herrera \textit{et al.} \cite{pin1, ya29,ya30}. They not only presented
the way to evaluate such scalar (orthogonal
splitting of Riemann tensor) but also used these variables in the modeling of the many stellar objects in the realm of GR. These are
\begin{equation}\label{19}
X_{\alpha\beta}=~^{*}R^{*}_{\alpha\gamma\beta\delta}V^{\gamma}V^{\delta}=
\frac{1}{2}\eta^{\varepsilon\rho}_{~~\alpha\gamma}R^{*}_{\epsilon
\rho\beta\delta}V^{\gamma}V^{\delta},\quad
Y_{\alpha\beta}=R_{\alpha\gamma\beta\delta}V^{\gamma}V^{\delta},
\end{equation}
where the notation steric on the both, left and right sides of the tensor
indicate double, left and right duals of the subsequent entities, respectively.
\begin{align}\nonumber
X_{\gamma\delta}&=X_{\gamma\delta}^{(m)}+X_{\gamma\delta}^{(D)}=\frac{1}{3}\left[
\bar{\mu}+\lambda\varepsilon-\frac{\alpha}{2}\left\{\beta(1-n-m)G^m+(1-n)\right\}G^n-\frac{\lambda T}{2}+\frac{\psi_{00}}{A^2
}\right]h_{\gamma\delta}\\\label{20a}
&-\frac{1}{2}\left[(1+\lambda)(\bar{\Pi}-2\eta\sigma)-\frac{{\psi}_{11}}{B^2}
+\frac{{\psi}_{22}}{C^2}\right]\left(\chi_\gamma\chi_\delta-\frac{1}{3}h_{\gamma\delta}\right)-E_{\gamma\delta},\\\nonumber
Y_{\gamma\delta}&=Y_{\gamma\delta}^{(m)}+Y_{\gamma\delta}^{(D)}=\frac{1}{6}\left[\bar{\mu}+\lambda\varepsilon+3\mu\lambda+(1+\lambda)
(3P_r-2\bar{\Pi})-\frac{\psi_{00}}{A^2}-\frac{\psi_{11}}{B^2}+\frac{2\psi_{22}}{C^2}+\frac{\alpha}{2}\right.\\\nonumber
&\left.\times\left\{\beta(1-n-m)G^m+(1-n)\right\}G^n+\frac{\lambda T}{2}\right]h_{\gamma\delta}+\frac{1}{2f_R}
\left[(1+\lambda)(\bar{\Pi}-2\eta\sigma)-\frac{{\psi}_{11}}{B^2}
+\frac{{\psi}_{22}}{C^2}\right]\\\label{20b}
&\times\left(\chi_\gamma\chi_\delta
-\frac{1}{3}h_{\gamma\delta}\right)-E_{\gamma\delta}.
\end{align}
Such tensorial forms can be given in another way via their trace (denoted with subscript T)and
trace-less (labeled with subscript TF) parts as
\begin{align}\label{19}
X_{\gamma\delta}&=\frac{1}{3}TrXh_{\gamma\delta}+X_{\langle\gamma\delta\rangle},\\\label{20}
Y_{\gamma\delta}&=\frac{1}{3}TrYh_{\gamma\delta}+Y_{\langle\gamma\delta\rangle},
\end{align}
where
\begin{align}\label{21}
X_{\langle\gamma\delta\rangle}&=h^\nu_\gamma
h^\mu_\delta\left(X_{\nu\mu}-\frac{1}{3}TrXh_{\nu\mu}\right),\\\label{22}
Y_{\langle\gamma\delta\rangle}&=h^\nu_\gamma
h^\mu_\delta\left(Y_{\nu\mu}-\frac{1}{3}TrYh_{\nu\mu}\right).
\end{align}
{}From Eqs.(\ref{18})-(\ref{20b}), we found
\begin{align}\nonumber
&TrX\equiv
X_{T}=\left\{\bar{\mu}+\lambda\varepsilon
-\frac{\alpha}{2}\left\{\beta(1-n-m)G^m+(1-n)\right\}G^n-\frac{\lambda T}{2}\right.\\\label{23}
&\left.-\frac{{\lambda}}{2}T
+\frac{\hat{\psi}_{00}}{A^2}\right\},\\\nonumber &TrY\equiv
Y_{T}=\left\{\bar{\mu}+\lambda\varepsilon+3\mu\lambda
+3(1+\lambda)\bar{P_r}-\frac{\hat{\psi}_{11}}{B^2}+2\alpha(1-n)R^n-\frac{\hat{\psi}_{00}}{A^2}\right.\\\label{24}
&\left.-2(1+\lambda)\bar{\Pi}-\frac{2\hat{\psi}_{22}}
{C^2}+2\beta(3-n) R^{2-n}-2{\lambda}T\right\},
\end{align}
where $X_{TF}$ and $Y_{TF}$ stand for the trace-free components of the
tensors $X_{\alpha\beta}$ and $Y_{\alpha\beta}$, respectively (for
details, please see \cite{ya31a,ya31b,ya31c,ya31d,ya31e,ya31f,ya31g,ya31h,ya31i}). We can also write
$X_{<\alpha\beta>}$ and $Y_{<\alpha\beta>}$ in an alternatively form
\begin{align}\label{25}
X_{\langle\gamma\delta\rangle}&=X_{TF}\left( \chi_{\gamma}\chi_{\delta}
-\frac{1}{3}h_{\gamma\delta}\right),\\\label{26}
Y_{\langle\gamma\delta\rangle}&=Y_{TF}\left( \chi_{\gamma}\chi_{\delta}
-\frac{1}{3}h_{\gamma\delta}\right),
\end{align}

Using Eqs.(\ref{6})-(\ref{10}), (\ref{20}) and (\ref{21}), we obtain
\begin{align}\label{23}
X_{TF}&=-\mathcal{E}-\frac{1}{2}
\left\{(\lambda +1)(-2{\sigma}{\eta}+\bar{\Pi})
+\frac{{\varphi}_{22}}{C^2}-\frac{{\varphi}_{11}}{H^2}\right\},\\\label{25}
Y_{TF}&=\mathcal{E}-\frac{1}{2}
\left\{(\bar{\Pi}-2{\eta}{\sigma})(\lambda +1)
+\frac{{\varphi}_{22}}{C^2}-\frac{{\varphi}_{11}}{H^2}\right\}.
\end{align}
The value of $Y_{TF}$ can be followed from Eqs.(\ref{18}) and
(\ref{25}) as
\begin{align}\nonumber
Y_{TF}&=\frac{1}{2}\left(\bar{\mu}+\varepsilon\lambda-\frac{\alpha}{2}\left\{\beta(1-m-n)G^m+(1-n)\right\}G^n-\frac{\lambda T}{2}-2(1+\lambda )(\bar
{\Pi}-4{\eta}{\sigma})\right.\\\nonumber
&\left.-\frac{{\varphi}_{00}}{A^2}+\frac{2{\varphi}_{11}}{H^2}
-\frac{2{\varphi}_{22}}{C^2}\right)-\frac{3}{2C^3}
\int^r_{0}\frac{C^2}{1+2R\lambda T^2}
\left[\bar{\mu}-\frac{\alpha}{2}\left\{\beta(1-m-n)G^m+(1-n)\right\}G^n \right.\\\label{26}
&\left.-\frac{\lambda T}{2}+\varepsilon\lambda+\frac{{\varphi}_{00}}{A^2}+\frac{U}{E}
\left\{(1+\lambda )\bar{q}
+\frac{{\varphi}_{01}}{AB}\right\}C^2C'\right]dr.
\end{align}

It can be interesting to visualize our set of equations in terms of some
dagger variables which are defined as under
\begin{align*}
\mu^{\dag}&\equiv\bar{\mu}-\frac{{\varphi}_{00}}{A^2}, \quad
P^{\dag}_{r}\equiv
\bar{P_r}-\frac{{\varphi}_{11}}{H^2}-\frac{4}{3}{\eta}{\sigma},\\
P^{\dag}_{\bot}&\equiv P_{\bot}-\frac{{\varphi}_{22}}{C^2}+\frac{2}{3}{\eta}\sigma,\\
\Pi^{\dag}&\equiv P^{\dag}_{r}-P^{\dag}_{\bot}=\Pi-2{\eta}{\sigma}
+\frac{\varphi_{22}}{C^2}-\frac{\varphi_{11}}{B^2}.
\end{align*}
Then, Eqs.(\ref{22})-(\ref{25}) gives
\begin{align}\nonumber
X_{TF}&=\frac{3}{2C^3} \int^r_{0}\left[\left\{\mu^{\dag}-\frac{\alpha}{2}\left\{G^m\beta(1-m-n)+(1-n)\right\}G^n-\frac{\lambda T}{2}+\lambda\varepsilon
+\left(\hat{q}(\lambda+1)+\frac{\varphi_q}{BA}\right)\right.\right.\\\label{27} &\left.\left.\times\frac{U}{E}\right
\} C^2C'\right]dr-\frac{1}{2}\left[\mu^{\dag}-\frac{\alpha}{2}\left\{G^m\beta(1-m-n)+(1-n)\right\}G^n-\frac{\lambda T}{2}   +\lambda\varepsilon\right],\\\nonumber
Y_{TF}&=\frac{1}{2}\left[\mu^{\dag}-\frac{\alpha}{2}\left\{G^m\beta(1-m-n)+(1-n)\right\}G^n-\frac{\lambda T}{2}+\varepsilon\lambda-2\lambda \right.\\\nonumber
&\left.\times\left(\frac{{\varphi}_{11}}{H^2}
-\frac{{\varphi}_{22}}{C^2}\right)\right]-\frac{3}{2C^3}\int^r_{0}\left[
\left\{\mu^{\dag}-\frac{\alpha}{2}\left\{G^m\beta(1-m-n)+(1-n)\right\}G^n-\frac{\lambda T}{2}+\lambda\varepsilon\right.\right.\\\label{28}
&\left.\left.+\left(\hat{q}(\lambda+1)+\frac{\varphi_q}
{BA}\right)\frac{U}{E}\right\}C^2C'\right]dr,\\\nonumber
Y_{T}&=\frac{1}{2}\left[(1+3\lambda)\mu^{\dag}
-2\lambda\varepsilon+3(1+\lambda)P_r^{\dag}-2\Pi^{\dag}(1+\lambda )
+\frac{\alpha}{2}\left\{\beta(1-m-n)G^m+(1-n)\right\}G^n\right.\\\label{29} &\left. +\lambda \left(2\frac{\varphi_{22}}{C^2}+\frac{\varphi_{11}}{B^2}+3
\frac{\varphi_{00}}{A^2}\right)+\frac{\lambda T}{2}\right],
\\\label{30}
X_{T}&=\mu^{\dag}-\frac{\alpha}{2}\left\{\beta G^m(1-m-n)+(1-n)\right\}G^n-\frac{\lambda T}{2}+\varepsilon\lambda.
\end{align}
These are $f(G,T)$ structure scalars which are four in number. These scalars occupy very important
role in the dynamical properties of the self-gravitating structures, for example,
irregular energy density, mass function, tidal forces, curvature of spacetime, etc.
The well-known equations like, shear evolution equation (SEE), expansion evolution equation (EEE) also known as Raychaudhuri equation
and the Weyl differential equation (WDE). The so called Raychaudhuri equation
was also calculated autonomously by Landau) \cite{ya36}.
Through $f(G,T)$ scalar variables, the one of the structure scalar $Y_T$ can be recasted as
\begin{equation}\label{31}
-(Y_{T})=\frac{1}{3}\left(2{\sigma}^{
\alpha\beta}{\sigma}_{\alpha\beta}+\Theta^{2}\right)+V^{\alpha}\Theta_{;\alpha}
-a^\alpha_{~;\alpha}.
\end{equation}
It can be checked from the above expression that the notable EEE can be well-written through one of the $f(G,T)$ matter
scalar. Similarly, the SEE can be manipulated through $Y_{TF}$ as follows
\begin{equation}\label{32}
\mathcal{E}-\frac{1}{2}
\left\{(\bar{\Pi}-2{\eta}{\sigma})(\lambda +1)
+\frac{{\varphi}_{22}}{C^2}-\frac{{\varphi}_{11}}{H^2}\right\}=Y_{TF}=a^{2}+\chi^{\alpha}a_{;\alpha}-\frac{aC'}{BC}
-\frac{1}{3}\left(2{\Theta}\sigma+\sigma^{2}\right)-V^\alpha
\sigma_{;\alpha}.
\end{equation}
Equations (\ref{16})-(\ref{16}) provides WDE for the shearing viscous spherical matter distribution as
\begin{align}\nonumber
&\left[X_{TF}+\frac{1}{2}\left(\mu^{\dag}-\frac{\alpha}{2}\left\{\beta(1-n-m)G^m+(1-n)\right\}G^n-\frac{\lambda T}{2}\right)\right]'=
-X_{TF}\frac{3C'}{C}+\frac{1}{2}(\Theta-\sigma),\\\label{33}
&\times\left({q}B(\lambda+1)+\frac{\varphi_q}{A}\right).
\end{align}
It can be seen from the above equation that in the configurations of WDE, the
$f(G,T)$ structure scalar, $X_{TF}$, has pivotal role. The solution of the above equation would present $X_{TF}$
as a factor of controlling inhomogeneous matter density in the background of relativistic spheres in $f(G,T)$ gravity.

\section{Dust Ball with Constant $G$ and $T$}

This section is devoted to examine the influences of $f(G)+\lambda
T$ MGT on the formulations of SEE, WDE and EEE for the pressure less
non-interacting particles with constant $G$ and $T$ terms. We shall use the notation tilde
to represent that the corresponding terms are compute with constant choices of $G$
and $T$. Thus, the mass function in the context turns out to be
\begin{align}\label{34}
\mathfrak{m}&=\frac{1}{2} \int^r_{0}\mu
C^2dC-\frac{{\lambda}R^2T^2} {2\{1+2R\lambda T^2\}}\int^r_{0} C^2C'dr.
\end{align}
The value of the Weyl scalar in an environment of relativistic dust ball is found to be
\begin{align}\label{35}
\mathcal{E}&=\frac{1}{2C^3}
\int^r_{0}\mu'C^3dr-\frac{\alpha}{2}\left\{\beta(1-n-m)G^m+(1-n)\right\}G^n-\frac{\lambda T}{2},
\end{align}
while the spherical dust ball mass the mass function
\begin{align}\label{36} \frac{3\mathfrak{m}}{C^3}&=\frac{1}{2}
\left[\mu-\frac{1}{C^3}\int^r_{0}\mu'C^3dr\right]-\frac{\alpha}{2}\left\{\beta G^m(1-m-n)+(1-n)\right\}G^n-\frac{\lambda T}{2}.
\end{align}
The set of scalar functions for the case of $f(G,T)$ gravity turns out to be
\begin{align}\label{37} \tilde{X}_{T}&={\mu}-\frac{\alpha}{2}\left\{\beta G^m(1-m-n)+(1-n)\right\}G^n-\frac{\lambda T}{2},\\\label{38}
\tilde{Y}_{TF}&=-\tilde{X}_{TF}=\mathcal{E},\\\label{39}
\tilde{Y}_{T}&=\frac{1}{2}\left[{\mu}+{\alpha}\left\{\beta(1-m-n)G^m+(1-n)\right\}G^n+{\lambda T}\right].
\end{align}
These expressions describe that the effects of the scalars, i.e., $Y_T,~X_T$ and $Y_{TF},~X_{TF}$ are
being controlled by matter variables like $f(G,T)$ terms, tidal forces and fluid energy density,
in the subsequent evolution of the star. The causes for the emergence of IED can be explored for the
dust matter through the following WDE as
\begin{align}\label{40}
&\left[\frac{\mu}{2}+
\frac{\alpha}{2}\left\{\beta G^m(1-m-n)+(1-n)\right\}G^n+\frac{\lambda T}{2}+\tilde{X}_{TF}\right]'
=-\frac{3}{C}\tilde{X}_{TF}C'.
\end{align}
This points out that the $f(G,T)$ scalar, i.e., $\tilde{X}_{TF}$ is a factor for producing and reducing inhomogeneities
over the surface of the initially homogeneous dust ball. One can analyze that $\mu=\mu(t)\Leftrightarrow\tilde{X}_{TF}=0=\alpha=\lambda$. This asserts that dark source terms coming from $f(G,T)$ gravity are trying to produce resistance against the fluctuations
of IED. Furthermore, the EEE and SEE boil down to
\begin{align}\nonumber
V^\alpha\Theta_{;\alpha}+\frac{2}{3}\sigma^2 +\frac{\Theta^2}{3}
-a^\alpha_{;\alpha}&=\frac{1}{2}\left[{\mu}+{\alpha}\left\{\beta(1-n-m)G^m+(1-n)\right\}G^n+{\lambda T}\right]\\\label{41}
&=-\tilde{Y}_T,\\\label{42}
V^{\alpha}\sigma_{;\alpha}+\frac{\sigma^{2}}{3}+\frac{2}{3}\sigma
\Theta&=-\mathcal{E}=-\tilde{Y}_{TF},
\end{align}
which shows the importance of $\tilde{Y}_{TF}$ and $\tilde{Y}_T$ in the definitions of
EEE and SEE.

\section{Conclusions}

In this paper, we have investigated the influences of $f(G,T)$
corrections on some dynamical properties of evolving stellar bodies.
For this purpose, we have considered spherically symmetric geometry
which is assumed to to coupled with anisotropic radiating matter
contents. We assumed that the relativistic fluid distribution has a
shearing viscous property and is emitting radiations in the free
streaming and diffusion approximations. The dissipation is carrying
out without scattering. We have then assumed extra degrees of
freedom mediated by $f(G,T)$ gravity. We have then calculated the
corresponding field equations and dynamical equation. The source of
producing tidal forces, i.e., Weyl scalar has been calculated and
then related it with the matter and metric variables along with dark
sources terms coming from polynomial $f(G,T)$ gravity. With the help
of notable Misner-Sharp mass function, we have calculated quantity
of matter content within the spherical geometry. We then expressed
such a relation with the previously calculated Weyl scalar equation.
This relation has peculiar importance in the modeling of stellar
structures.

We have then broke down the expression of the Riemann curvature
tensor by applying the orthogonal decomposition technique. The
technique that was firstly developed by Herrera \emph{et al.} \cite{pin1}.
We have applied this technique for our stellar model with one of the
modified gravity theories, i.e., $f(G,T)$ gravity. With this, we
have developed relations of two tensor namely, $X_{\mu\nu}$ and
$Y_{\mu\nu}$. The trace and trace-free parts of these tensorial
quantities are computed, which have utmost relevance in the study of
gravitational collapse, stellar evolution, etc. These trace and
trace-less parts are called here as $f(G,T)$ structure scalars. We
have then evaluated a well-known Raychaudhuri equation and expressed
it in terms of these structure scalars. It is worthy to mention that
this equation play a vital role in the discussion of the
Penrose-Hawking singularity theorems. This equation has also been
used to find many exact solutions of gravitational field equations
in literature. Thus, our one of the structure scalar $Y_T$ occupy
fundamental importance in understanding the scenario under which
gravitational effects could lead to singularities. Thus, $Y_T$ could be
helpful to understand the singularity appearances in various black
holes, like Schwarzschild, Kerr, the Reissner–Nordstr\"{o}m and the
Kerr-Newman metrics etc.

After that, we have evaluated shear and Weyl scalar evolution
equations through $f(G,T)$ structure scalars. The Weyl
scalar equation that is expressed through $X_{TF}$ describes the
propagation of tidal forces in the modeling of radiating spherical
stars in modified gravity. This scalar has utmost relevance in the
study of Newman-Penrose formalism along with degrees of freedom
mediated by $f(G,T)$ gravity. The scalar $X_{TF}$ which is related with
Weyl scalar could be fruitful to examine the outgoing gravitational
radiation from the asymptotically flat geometry. Finally, we have
encode all of our results for the case of non-dissipative dust ball
with constant values of $G$ and $T$. In this scenario, we have shown
that two $f(G,T)$ scalar functions are directly related with Weyl
scalar and thus tidal forces, Further, the $f(G,T)$ scalar $X_{TF}$
is controlling the appearance of inhomogeneities in the initially
regular compact body.

\vspace{0.25cm}

{\bf Acknowledgments}

\vspace{0.25cm}

This work is supported by National Research Project for Universities
(NRPU), Higher Education Commission, Islamabad under the research project
No. 8754.


\vspace{0.5cm}

\begin{thebibliography}{40}

\bibitem{qs1} E. Berti et al., Class. Quantum Grav. \textbf{32}, 243001 (2015).

\bibitem{qs2} A. De Felice and S. Tsujikawa, Living Rev. Relat. \textbf{13}, 3 (2010) [arXiv:1002.4928 [gr-qc]].

\bibitem{qs3} S. Capozziello and M. De Laurentis, Phys. Rep. \textbf{509}, 167 (2011) [arXiv:1108.6266 [gr-qc]].

\bibitem{qs4} S. Nojiri and S. D. Odintsov, Phys. Rep. \textbf{505}, 59 (2011) [arXiv:1011.0544 [gr-qc]].

\bibitem{qs5} P. Bull et al., Phys. Dark Universe, \textbf{12}, 56 (2016) [arXiv:1512.05356 [astro-ph.CO]].

\bibitem{qs6} L. Amendola et al., Living Rev. Relat. \textbf{16}, 6 (2013) [arXiv:1206.1225 [astro-ph.CO]].

\bibitem{qs7} L. Amendola et al., arXiv:1606.00180 [astro-ph.CO].

\bibitem{zs3} A. Qadir, H. W. Lee, and K. Y. Kim, Int. J. Mod. Phys. D \textbf{26}, 1741001 (2017).

\bibitem{v41} T. P. Sotirou and V. Faraoni, Rev. Mod. Phys. \textbf{82}, 451 (2010).

\bibitem{b2a} S. Capozziello and M. D. Laurentis, Phys. Rep. \textbf{509}, 167 (2011).

\bibitem{R-DE-MG} A. D. Felice and S. Tsujikawa, Living Rev. Relativ. \textbf{13}, 3
(2010).

\bibitem{martin1} S. Nojiri and S. D. Odintsov, Phys. Rep. \textbf{505}, 59 (2011) [arXiv:1011.0544
[gr-qc]].

\bibitem{ya3} S. Nojiri and S. D. Odintsov,
eConf C {\bf 0602061}, 06 (2006) [Int. J. Geom. Meth. Mod. Phys.
\textbf{4}, (2007) 115] [hep-th/0601213].

\bibitem{mnras1} Z. Yousaf and M. Z. Bhatti, Mon. Not. R. Astron. Soc. \textbf{458}, 1785 (2016) [arXiv:1612.02325 [physics.gen-ph]].

\bibitem{martin3} S. Nojiri and S. D. Odintsov, Phys. Lett. B \textbf{631}, 1 (2005).

\bibitem{g11} S. Nojiri, S. D. Odintsov and S. Ogushi, Int. J. Mod. Phys. A \textbf{17}, 4809 (2002).
\bibitem{g12} B. M. Leith and I. P. Neupane, J. Cosmol. Astropart. Phys. \textbf{0705}, 019 (2007).

\bibitem{ya9} T. Harko, F. S. N. Lobo, S. Nojiri and S. D. Odintsov, Phys. Rev. D
\textbf{84}, 024020 (2011).

\bibitem{ya10} M. J. S. Houndjo, Int. J. Mod. Phys. D \textbf{21}, 1250003 (2012).

\bibitem{ya13} E. H. Baffou, A. V. Kpadonou, M. E. Rodrigues,
M. J. S. Houndjo and J. Tossa, Astrophys. Space Sci. \textbf{356},
173 (2014).

\bibitem{z5d} K.~Bamba, S.~Capozziello, S.~Nojiri and S.~D.~Odintsov, Astrophys.\
Space Sci. {\bf 342}, 155 (2012).

\bibitem{mart5} S. Capozziello, M. De Laurentis, I. De
Martino, M. Formisano and S. D. Odintsov, Phys. Rev. D \textbf{85}
044022, (2012) [arXiv: 1112.0761].

\bibitem{mart6} S. Capozziello, M. De Laurentis, S. D. Odintsov and A. Stabile Phys.
Rev. D \textbf{83}, 064004 (2011) [arXiv: 1101.0219 [gr-qc]].

\bibitem{mart7} K. Bamba, S. Nojiri and S. D. Odintsov, Phys. Lett. B \textbf{698},
451 (2011) [arXiv: 1101.2820].

\bibitem{7p1} M. Sharif and Z. Yousaf, Eur. Phys. J. C \textbf{75}, 58 (2015).

\bibitem{7p2} M.~Z.~Bhatti,~Z.~Yousaf~and~S.~Hanif,~Mod.~Phys.~Lett.~A~\textbf{32},~1750042~(2017).

\bibitem{7p3} M.~Z.~Bhatti,~Z.~Yousaf~and~S.~Hanif, Phys. Dark Universe \textbf{75}, 58 (2017).

\bibitem{7s1} M. Sharif and Z. Yousaf, Eur. Phys. J. C  \textbf{75}, 194 (2015) [arXiv:1504.04367v1 [gr-qc]].

\bibitem{7s2} M. Sharif and Z. Yousaf, Astrophys. Space Sci.  \textbf{355}, 317 (2015).

\bibitem{7s3} Z. Yousaf, K. Bamba and M. Z. Bhatti, Phys. Rev. D \textbf{93}, 064059 (2016) [arXiv1603.03175 [gr-qc]].

\bibitem{7s4} Z. Yousaf, Astrophys. Space Sci. \textbf{363}, 226 (2018).

\bibitem{7s5} Z. Yousaf, Eur. Phys. J. Plus \textbf{132}, 71 (2017).

\bibitem{7s6} Z.~Yousaf,~M.~Z.~Bhatti and A. Rafaqat,~Astrophys. Space Sci. \textbf{68}, 362 (2017).

\bibitem{7c1} M. Sharif and Z. Yousaf, Astrophys. Space Sci. \textbf{357}, 49 (2015).

\bibitem{7c3} Z. Yousaf and M. Z.
Bhatti,  Eur. Phys. J. C \textbf{76}, 267 (2016) [arXiv:1604.06271
[physics.gen-ph]].

\bibitem{9} P. H. R. S. Moraes, J. D. V. Arba\~{n}il and M. Malheiro, J. Cosmol.
Astropart. Phys. \textbf{06}, 005 (2016).

\bibitem{gt1} M. Z. Bhatti, M. Sharif, Z. Yousaf and M. Ilyas, Int. J. Mod. Phys. D \textbf{27}, 1850044 (2018).

\bibitem{12} L. Herrera, A. Di Prisco, G. Le Denmat, M. A. H. MacCallum and N. O. Santos,
Phys. Rev. D \textbf{76}, 064017 (2007).

\bibitem{13} L. Herrera and N. O. Santos, Class. Quantum Grav. \textbf{22}, 2407 (2005).

\bibitem{15} B. C. Tewari, K. Charan and J. Rani, Int. J. Astron. Astrophys. \textbf{6}, 155 (2016).

\bibitem{16rt} M. Sharif and Z. Yousaf, Astrophys. Space Sci. \textbf{354}, 471 (2014).

\bibitem{16r} M. Sharif and Z. Yousaf, Int. J. Theor. Phys. \textbf{55}, 470 (2016).

\bibitem{epjc1} Z. Yousaf, M. Z. Bhatti and U. Farwa, Eur. Phys. J. C \textbf{77}, 359
(2017).

\bibitem{cqg} Z. Yousaf, M. Z. Bhatti and U. Farwa, Class. Quantum Grav. \textbf{34}, 145002
(2017).

\bibitem{sah1} P. K. Sahoo, P. Sahoo and B. K. Bishi, Int. J. Geom. Meth. Mod. Phys. \textbf{14}, 1750097 (2017).

\bibitem{sah2a} P. H. R. S. Moraes and P. K. Sahoo, Eur. Phys. J. C \textbf{77}, 480 (2017).

\bibitem{sah2b} P. H. R. S. Moraes and P. K. Sahoo, Phys. Rev. D \textbf{96}, 044038 (2017).

\bibitem{sah2c} P. H. R. S. Moraes,
P. K. Sahoo, G. Ribeiro and R. A. C. Correa, arXiv:1712.07569 [gr-qc].

\bibitem{ya26} R. Penrose and S. W. Hawking, \emph{General Relativity, An Einstein Centenary
Survey}, Cambridge University Press, Cambridge (1979).

\bibitem{ya27} L. Herrera, A. Di Prisco, J. L. Hern\'{a}ndez-Pastora and N. O. Santos,
Phys. Lett. A \textbf{237}, 113 (1998).

\bibitem{vir1a} K. S. Virbhadra, D. Narasimha, and S. M. Chitre, Astron.
Astrophys. \textbf{337}, 1 (1998).

\bibitem{vir1b} K. S. Virbhadra and G. F. R. Ellis, Phys. Rev. D \textbf{65}, 103004 (2002).

\bibitem{vir2} K. S. Virbhadra, Phys. Rev. D \textbf{79}, 083004 (2009).

\bibitem{ya29} L. Herrera, A. Di Prisco, J. Martin, J. Ospino, N. O. Santos and O. Troconis, Phys. Rev. D \textbf{69}, 084026 (2004).

\bibitem{ya30} L. Herrera, A. Di Prisco and J. Ib\'{a}\~{n}ez, Phys. Rev. D \textbf{84}, 107501 (2011).

\bibitem{y1t} Z.~Yousaf,~K.~Bamba~and~M.~Z.~Bhatti,~Phys. Rev. D \textbf{93}, 124048 (2016) [arXiv:1606.00147 [gr-qc]].

\bibitem{b1ta} M. Z. Bhatti and Z. Yousaf, Eur. Phys. J. C \textbf{76}, 219 (2016) [arXiv1604.01395 [gr-qc]].

\bibitem{b1tb}  M. Z. Bhatti and Z. Yousaf, Int. J. Mod. Phys. D \textbf{26}, 1750029 (2017).

\bibitem{ltb} L. Herrera, A. Di Prisco and J. Iba\~{n}ez, Phys. Rev. D \textbf{84}, 064036 (2011).

\bibitem{entropy1} L. Herrera, Entropy \textbf{19}, 110 (2017) [arXiv:1703.03958 [gr-qc]].

\bibitem{y2t} Z.~Yousaf,~K.~Bamba~and~M.~Z.~Bhatti,~Phys. Rev. D \textbf{95}, 024024 (2017) [arXiv:1701.03067 [gr-qc]].

\bibitem{jpc} L. Herrera, J. Phys. Conf. Ser. \textbf{831}, 012001 (2017) [arXiv:1704.04386 [gr-qc]].

\bibitem{g16} K. Bamba, S. D. Odintsov, L. Sebastiani, S. and Zerbini, Eur. Phys. J. C \textbf{67}, 295 (2010).

\bibitem{ya32} C. W. Misner and D. Sharp, Phys. Rev. \textbf{136}, B571 (1964).

\bibitem{ya35} L. Bel, Ann. Inst. H Poincar\'{e} \textbf{17}, 37 (1961).

\bibitem{pin1} L. Herrera, J. Ospino, A. Di Prisco, E. Fuenmayor, O. Troconis, Phys. Rev. D \textbf{79}, 064025 (2009).

\bibitem{ya31a} M. Sharif and M. Z. Bhatti, Int. J. Mod. Phys. D
\textbf{23}, 1450085 (2014).

\bibitem{ya31b} M. Sharif and M. Z. Bhatti, \textbf{24}, 1550014 (2015).

\bibitem{ya31c} M. Sharif and M. Z. Bhatti, Mod. Phys. Lett. A \textbf{29}, 1450165 (2014).

\bibitem{ya31d} M. Sharif and M. Z. Bhatti, \textbf{29}, 1450094
(2014).

\bibitem{ya31e} M. Sharif and Z. Yousaf, Astrophys. Space Sci. \textbf{352},
321 (2015).

\bibitem{ya31f} M. Sharif and Z. Yousaf, Astrophys. Space Sci. \textbf{354}, 481 (2014).

\bibitem{ya31g} M. Sharif and Z. Yousaf, Astrophys. Space Sci. \textbf{357}, 49
(2015).

\bibitem{ya31h} M. Sharif and Z. Yousaf, Eur. Phys. J. C \textbf{75}, 194 (2015).

\bibitem{ya31i} M. Sharif and Z. Yousaf, Can. J. Phys. \textbf{93}, 905 (2015).

\bibitem{ya36} F. D. Albareti, J. A. R. Cembranos, \'{A}. de la Cruz-Dombriz and A. Dobado, J.
Cosmol. Astropart. Phys. \textbf{03}, 012 (2014).

\end{thebibliography}
\end{document}